\documentclass{pasj00}

\def\submitted{submitted}
\def\inpress{in press}

\def\arxiv#1{ (arXiv astro-ph/#1)}

\DeclareAbbreviation\AAHam{Astron. Abh. Hamburg. Sternw.}
\DeclareAbbreviation\AARv{Astron. Astrophys. Rev.}
\DeclareAbbreviation\AAS{American Astron. Soc. Meeting Abstracts}
\DeclareAbbreviation\AcA{Acta Astron.}
\DeclareAbbreviation\actaa{Acta Astron.}
\DeclareAbbreviation\Afz{Astrofizika}
\DeclareAbbreviation\AGAb{Astronomische Gesellschaft Abstract Ser.}
\DeclareAbbreviation\an{Astron. Nachr.}
\DeclareAbbreviation\AnAp{Annales d'Astrophysique}
\DeclareAbbreviation\AnTok{Tokyo Astron. Obs. Annals, Sec. Ser.}
\DeclareAbbreviation\Ap{Astrophysics}
\DeclareAbbreviation\ARep{Astron. Rep.}
\DeclareAbbreviation\AstBu{Astrophys. Bull.}
\DeclareAbbreviation\ATel{Astron. Telegram}
\DeclareAbbreviation\ATsir{Astron. Tsirk.}
\DeclareAbbreviation\AcApS{Acta Astrophys. Sinica}
\DeclareAbbreviation\AstL{Astron. Lett.}
\DeclareAbbreviation\BaltA{Baltic Astron.}
\DeclareAbbreviation\BANS{Bull. of the Astron. Institutes of the Netherlands Suppl. Ser.}
\DeclareAbbreviation\BASI{Bull. Astron. Soc. India}
\DeclareAbbreviation\BeSN{Be Newslett.}
\DeclareAbbreviation\BHarO{Harvard Coll. Obs. Bull.}
\DeclareAbbreviation\CBET{Cent. Bur. Electron. Telegrams}
\DeclareAbbreviation\ChJAA{Chinese J. of Astron. and Astrophys.}
\DeclareAbbreviation\caa{Chinese J. of Astron. and Astrophys.}
\DeclareAbbreviation\CoAsi{Asiago Contr.}
\DeclareAbbreviation\CoSka{Contributions of the Astronomical Observatory Skalnat\'e Pleso}
\DeclareAbbreviation\GCN{GRB Coord. Netw. Circ.}
\DeclareAbbreviation\ErgAN{Erg. Astron. Nachr.}
\DeclareAbbreviation\ibvs{IBVS}
\DeclareAbbreviation\IEEEP{IEEE Proc.}
\DeclareAbbreviation\JAD{J. Astron. Data}
\DeclareAbbreviation\JApA{J. of Astrophys. and Astron.}
\DeclareAbbreviation\JAVSO{J. American Assoc. Variable Star Obs.}
\DeclareAbbreviation\JBAA{J. Br. Astron. Assoc.}
\DeclareAbbreviation\JPhCS{J. of Physics Conference Series}
\DeclareAbbreviation\JPSJ{J. Phys. Soc. Japan}
\DeclareAbbreviation\JSARA{J. of the Southeastern Assoc. for Research in Astron.}
\DeclareAbbreviation\LowOB{Lowell Obs. Bull.}
\DeclareAbbreviation\MitAG{Mitteil. der Astronom. Gesell. Hamburg}
\DeclareAbbreviation\MitVS{Mitteil. Ver\"{a}nderl. Sterne}
\DeclareAbbreviation\MmSAI{Mem. Soc. Astron. Ital.}
\DeclareAbbreviation\memsai{Mem. Soc. Astron. Ital.}
\DeclareAbbreviation\Msngr{Messenger}
\DeclareAbbreviation\NewA{New Astron.}
\DeclareAbbreviation\na{New Astron.}
\DeclareAbbreviation\NewAR{New Astron. Rev.}
\DeclareAbbreviation\nar{New Astron. Rev.}
\DeclareAbbreviation\NInfo{Nauchnye Informatsii}
\DeclareAbbreviation\NPhS{Nature Physical Science}
\DeclareAbbreviation\OAP{Odessa Astron. Publ.}
\DeclareAbbreviation\Obs{Observatory}
\DeclareAbbreviation\OEJV{Open Eur. J. on Variable Stars}
\DeclareAbbreviation\PASA{Publ. Astron. Soc. Australia}
\DeclareAbbreviation\PASAu{Publ. Astron. Soc. Australia}
\DeclareAbbreviation\PAZh{Pis'ma AZh}
\DeclareAbbreviation\PJAB{Proc. Japan Acad. Ser. B}
\DeclareAbbreviation\POBeo{Publ. de l'Observatoire Astronomique de Beograd}
\DeclareAbbreviation\PCCP{Phys. Chem. Chem. Phys.}
\DeclareAbbreviation\PhR{Phys. Rep.}
\DeclareAbbreviation\PVSS{Publ. Variable Stars Sect. R. Astron. Soc. New Zealand}
\DeclareAbbreviation\PZ{Perem. Zvezdy}
\DeclareAbbreviation\PZP{Perem. Zvezdy, Prilozh.}
\DeclareAbbreviation\QJRAS{QJRAS}
\DeclareAbbreviation\RA{Ricerche Astronomiche}
\DeclareAbbreviation\RMxAA{Rev. Mexicana Astron. Astrof.}
\DeclareAbbreviation\RvMA{Reviews of Modern Astron.}
\DeclareAbbreviation\SASS{Society for Astronom. Sciences Ann. Symp.}
\DeclareAbbreviation\Sci{Science}
\DeclareAbbreviation\SPIE{SPIE Proc.}
\DeclareAbbreviation\SvA{Soviet Astronomy}
\DeclareAbbreviation\SvAL{Soviet Astronomy Letters}
\DeclareAbbreviation\VeSon{Ver\"{o}ff. Sternw. Sonneberg}
\DeclareAbbreviation\VSOLJBul{VSOLJ Variable Star Bull.}
\DeclareAbbreviation\yCat{VizieR Online Data Catalog}
\DeclareAbbreviation\ZA{Z. Astrophys.}

\def\PublisherCambridge{Cambridge: Cambridge University Press}

\newcounter{author}
\def\authorcount#1#2{\refstepcounter{author}\label{#1}
                     \altaffiltext{\ref{#1}}{#2}}

\begin{document}
\SetRunningHead{T. Kato et al.}{Double Superoutburst in SSS J122221.7$-$311523}

\Received{201X/XX/XX}%{yyyy/mm/dd}
\Accepted{201X/XX/XX}%{yyyy/mm/dd}

\title{SSS J122221.7$-$311523: Double Superoutburst in a Best Candidate Period Bouncer}

\author{Taichi~\textsc{Kato},\altaffilmark{\ref{affil:Kyoto}*}
        Berto~\textsc{Monard},\altaffilmark{\ref{affil:Monard2}}
        Franz-Josef~\textsc{Hambsch},\altaffilmark{\ref{affil:Hambsch}}
        Seiichiro~\textsc{Kiyota},\altaffilmark{\ref{affil:Kis}}
        Hiroyuki~\textsc{Maehara},\altaffilmark{\ref{affil:Kiso}}
}

\authorcount{affil:Kyoto}{
     Department of Astronomy, Kyoto University, Kyoto 606-8502}
\email{$^*$tkato@kusastro.kyoto-u.ac.jp}

\authorcount{affil:Monard2}{
     Kleinkaroo Observatory, Center for Backyard Astronomy Kleinkaroo,
     Sint Helena 1B, PO Box 281, Calitzdorp 6660, South Africa}

\authorcount{affil:Hambsch}{
     Vereniging Voor Sterrenkunde (VVS), Oude Bleken 12, 2400 Mol, Belgium}

\authorcount{affil:Kis}{
     Variable Star Observers League in Japan (VSOLJ), 405-1003 Matsushiro,
     Tsukuba, Ibaraki 305-0035}

\authorcount{affil:Kiso}{
     Kiso Observatory, Institute of Astronomy, School of Science, 
     The University of Tokyo 10762-30, Mitake, Kiso-machi, Kiso-gun,
     Nagano 397-0101}

%%% end:list of authors


\maketitle

\begin{abstract}
We observed the 2012--2013 superoutburst of the newly identified
transient SSS J122221.7$-$311523 and found that this object
showed successive two superoutbursts.  Superhumps grew in amplitude
during the second superoutburst and showed a characteristic pattern
of period change reflecting the growth of the superhump.
Assuming that the periods of superhumps during the growing stage
[0.07721(1)~d] and post-superoutburst stage [0.07673(3)~d],
represent the dynamical precession rates at the radius of the 3:1
resonance and the radius immediately after the superoutburst,
respectively, we found that this object has a very small mass ratio  
$q=M_2/M_1 < 0.05$.  The possible orbital period from quiescent
data suggests $q=0.045$, one of the smallest among hydrogen-rich
cataclysmic variables.  The long orbital period and low $q$
make this object a perfect candidate for a period bouncer.
We suggest that the peculiar pattern of double superoutburst
is a result of a low $q$ and may be characteristic to
period bouncers.
\end{abstract}

\section{Introduction}

   Cataclysmic variables (CVs) are close binary systems
consisting of a white dwarf and a red-dwarf secondary 
transferring matter via the Roche-lobe overflow
[for a review, see \citet{war95book}].  According to the standard
scenario of CV evolution, CVs with longer orbital periods
($P_{\rm orb}$) evolve toward shorter $P_{\rm orb}$ due to
angular momentum loss due to the process of magnetic braking and
gravitational wave radiation.  At a certain point (called
the period minimum), the thermal time-scale of the secondary exceeds 
the mass-transfer time-scale and the mass-radius relation is 
reversed for degenerate dwarfs, $P_{\rm orb}$ starts to
lengthen (e.g. \cite{kol99CVperiodminimum}).  These objects
are usually called period bouncers.  Although the theory
predicts the majority of CVs have already evolved beyond
the period minimum, the population appears to be smaller 
than expected \citep{gan09SDSSCVs} and these objects are
observationally still elusive \citep{pat11CVdistance}.
Using the new technique based on the dynamical precession rate of
the growing superhumps (\cite{osa13v344lyrv1504cyg};
\cite{kat13qfromstageA}), we report on the detection of 
a dwarf nova system with one of the smallest
mass ratios ($q=M_2/M_1$) among hydrogen-rich CVs, hence
one of the best candidates for true period bouncers,
and discuss the interpretation of its peculiar superoutburst.

\section{SSS J122221.7$-$311523}

\begin{figure*}
  \begin{center}
%    \FigureFile(92mm,80mm){j1222lc.eps}
    \FigureFile(150mm,100mm){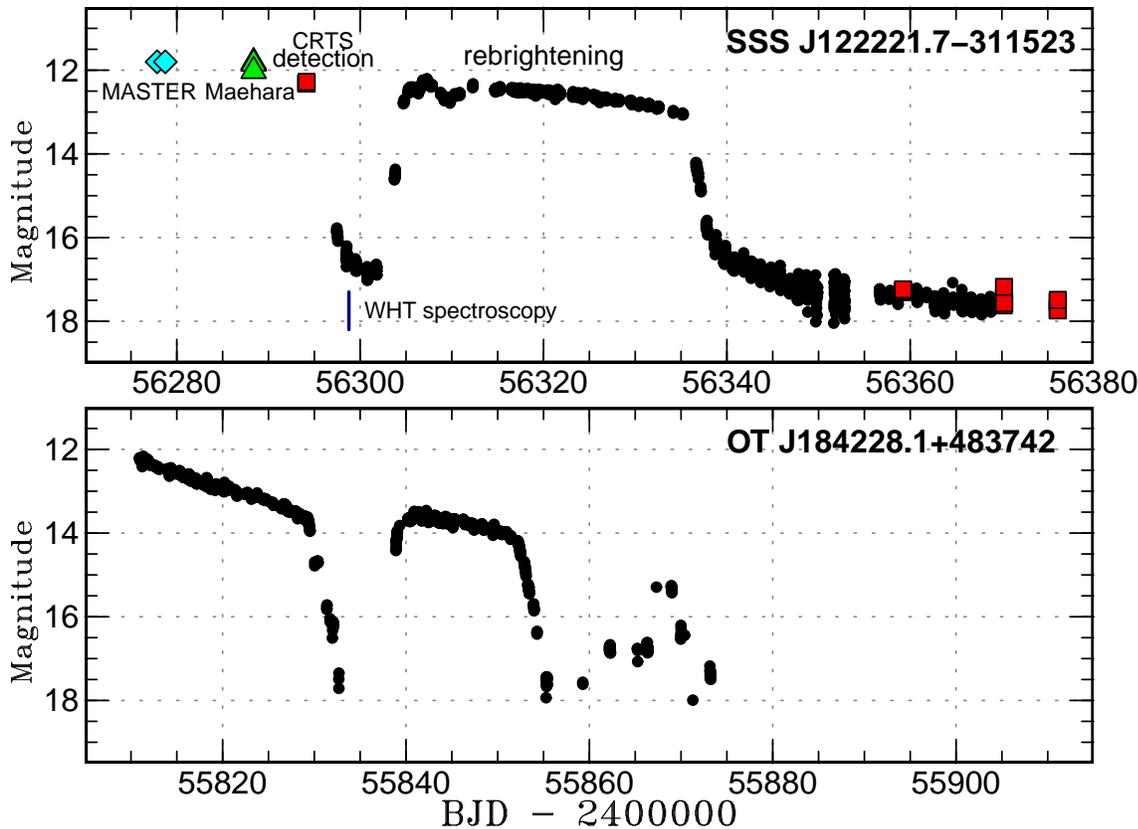}
  \end{center}
  \caption{Light curves of SSS J122221 and the similar
  period bouncer candidate OT J184228.
  The data for OT J184228 were taken from \citet{Pdot4}.
  The filled circles represents our observations (binned to 0.01~d).
  The symbols for the SSS J122221 are: CRTS SSS data (filled squares),
  Maehara Kyoto and Kiso Wide-field Survey data (filled triangles)
  and MASTER network data (filled diamonds).  The observations were
  mainly performed with unfiltered CCDs, whose magnitude system is
  close to $V$ for outbursting dwarf novae.}
  \label{fig:j1222lc}
\end{figure*}

   SSS J122221.7$-$311523 is a transient discovered
by Catalina Real-time Transient Survey (CRTS, \cite{CRTS})
Siding Spring Survey (SSS) (=SSS130101:122222-311525, 
hereafter SSS J122221)
on 2013 January 1 at $V=12.3$ \citep{dra13j1222atel4699}.
No secure previous outbursts were recorded in ASAS-3 data
(\cite{ASAS3}, 2000 November--2009 August, 627 nights)\footnote{
  There was one possible detection at $V$=14.56 on 2006 May 1.
  We regard it likely a noise since the object was recorded
  in usual quiescence three days later in the CRTS SSS data.
} and CRTS SSS data
(2005 August--2012 August, 79 nights), and the outburst amplitude
of $\sim$7 mag appeared to qualify the object to be 
a WZ Sge-type dwarf nova [for WZ Sge-type dwarf novae, 
see \citet{bai79wzsge}; \citet{kat01hvvir}; \citet{Pdot}].
H. Maehara reported that this object was already in outburst 
on 2012 December 26 at $V$=11.80--12.02 (vsnet-alert 15240).\footnote{
VSNET-alert archive can be accessed at
$<$http://ooruri.kusastro.kyoto-u.ac.jp/pipermail/vsnet-alert/$>$.
}
\citet{lev13j1222atel4700} reported that MASTER network
recorded that this object at 11.8 mag (unfiltered
CCD magnitude) on December 16.357 and 17.248 UT.  The object
must have been in a bright state for more than 16~d.

   \citet{mar13j1222atel4704} reported WHT spectroscopy taken on
2013 January 3, when double-peaked Balmer emission
lines were observed.  Although \citet{mar13j1222atel4704} noticed
the absence of an absorption component and the weakness of
the high-ionization lines which are unusual for an outbursting
dwarf nova, the object may have already faded from the outburst
at the time of the observation by \citet{mar13j1222atel4704},
since the object already faded to a magnitude of 15.5 on 2013
January 5 (vsnet-alert 15248).

   \citet{kuu13j1222atel4716} conducted a Swift target-of-opportunity
observation and reported optical spectroscopy on January 6,
yielding a preliminary orbital period of 80--95 min.
We should note that these observations were performed
during the temporary fading (see the mark in figure \ref{fig:j1222lc})
between the first outburst and the next outburst, 
which we call a rebrightening.
As we will see later, this rebrightening bears characteristics
of a superoutburst, and should be regarded distinct from
short rebrightening(s) seen in many SU UMa-type dwarf novae.

   The object remained faint until January 9.  On January 11,
the object started to brighten again rather slowly
(vsnet-alert 15262, 15269).
This phenomenon was also reported by \citet{neu13j1222atel4744}.
There was a precursor-like structure in the early part
of the rebrightening, and reached a local minimum on
January 16--17 (BJD 2456309--2456310).

   The object entered the rapid fading phase on February 12--13
(vsnet-alert 15385).  The object remained brighter (16--17 mag)
than in quiescence after this fading (see the upper panel of
figure \ref{fig:j1222lc} for the outburst light curve).

\section{Observation and Analysis}\label{sec:obs}

   The data were acquired by time-resolved unfiltered CCD photometry
(table \ref{tab:obs}).
All the observed times were corrected to barycentric Julian days (BJD).
Before making the analysis, we corrected zero-point differences
between different observers by adding a constant to each observer.
The data analysis was performed just in the same way described
in \citet{Pdot} and \citet{Pdot3}.

   In making period analysis, we used the phase dispersion minimization
(PDM) method \citep{PDM}.  We subtracted the global trend
of the outburst light curve by subtracting a smoothed light
curve obtained by low-order (up to 3) polynomials for
1--33~d segments (depending on the complexity of the light
curve) before the PDM analysis.
The 1$\sigma$ error of the PDM analysis was determined
by the methods in \citet{fer89error} for the Lafler-Kinman-type
period estimation.

\begin{table}
\caption{Summary of time-resolved observations.}\label{tab:obs}
\begin{center}
\begin{tabular}{cl}
\hline
Observer & Dates (BJD$-$2456000, number of observations \\
(telescope) & in the parentheses). \\
\hline
Monard & 297(412), 298(457), 299(551), 305(607), 306(542), \\
(35 cm) & 316(487), 317(417), 319(1098), 320(1029), \\
       & 321(1239), 324(1102), 325(1092), 329(1121), \\
       & 330(1358), 331(1216), 332(1255) \\
Hambsch & 301(44), 302(45), 304(43), 305(46), 306(51), \\
(40 cm) & 307(42), 308(75), 309(69), 310(45), 311(56), \\
       & 315(108), 317(68), 318(116), 319(119), 320(62), \\
       & 321(45), 322(46), 324(112), 326(218), 327(115), \\
       & 328(94), 337(124), 338(73), 339(116), 340(106), \\
       & 341(109), 342(50), 343(72), 344(121), 345(100), \\
       & 346(65), 347(34), 348(99), 349(116), 350(122) \\
Kiyota & 305(293), 307(295), 309(336), 310(220), 312(106), \\
(20 cm) & 315(284), 317(310), 318(281), 319(320), 321(283), \\
       & 323(246), 324(293), 326(228), 328(91), 334(122), \\
       & 335(244), 337(177) \\
\hline
\end{tabular}
\end{center}
\end{table}

\section{Results}\label{sec:result}

\subsection{Superhumps}\label{sec:sh}

   Despite the expectation as a WZ Sge-type dwarf nova,
the superhump signal was not clearly detected.
It finally started to grow following the start of 
the rebrightening (vsnet-alert 15275).  The true period was
identified 10~d after the rise to the rebrightening, and was 
unexpectedly long (longer than 0.07~d, which is well above
the usual range of WZ Sge-type dwarf novae; 
vsnet-alert 15302, 15306).

\begin{figure*}
  \begin{center}
%    \FigureFile(91mm,71mm){j1222humpall.eps}
    \FigureFile(150mm,100mm){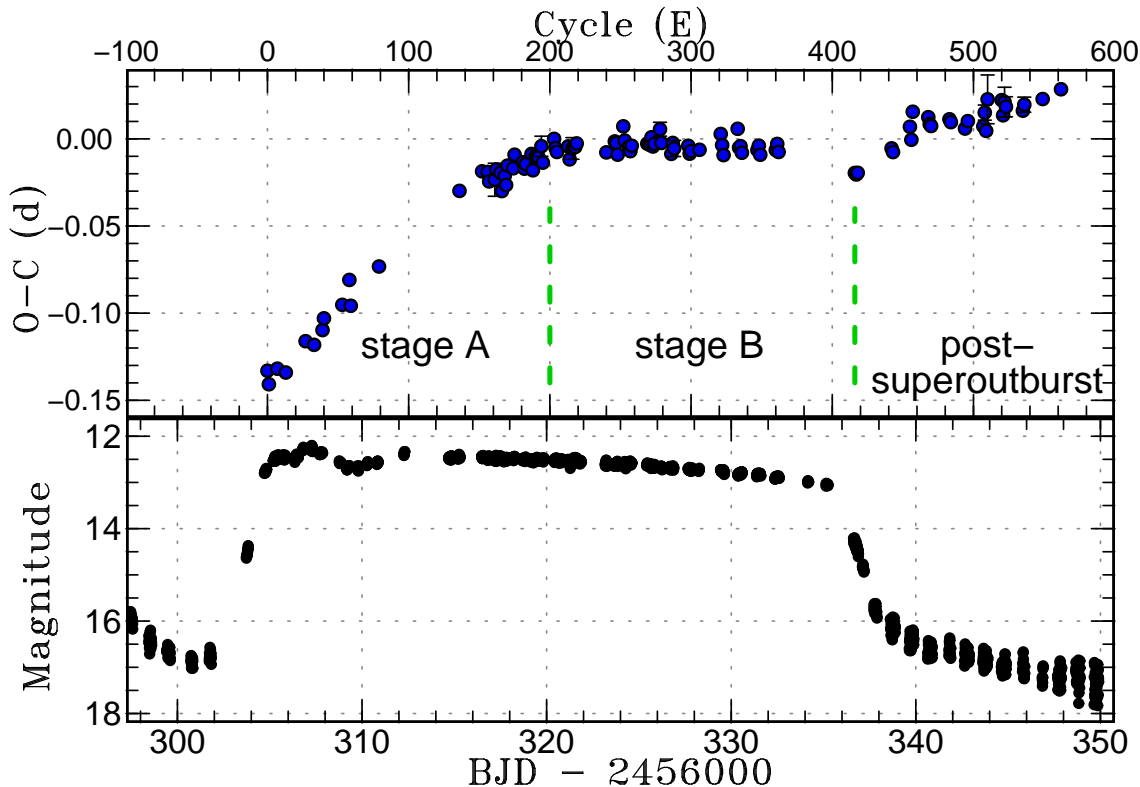}
  \end{center}
  \caption{$O-C$ variation of superhumps during the rebrightening.
  (Upper:) $O-C$.  The figure was drawn against the period of
  0.07649~d.
  (Lower:) Light curve (binned to 0.0076~d).}
  \label{fig:j1222humpall}
\end{figure*}

   By identifying the true superhump period, we have been able
to trace back the cycle counts and identified that the superhump
period was longer during the earlier half of the rebrightening
phase.  The times of superhumps maxima and other information
will be listed in our next summary paper on SU UMa-type 
dwarf novae (Kato et al. in prep.).
The pattern of period variation
looks unusual for a WZ Sge-type dwarf nova in that the period
showed a systematic decrease (figure \ref{fig:j1222humpall}),
while the superhump period usually increase in most of
WZ Sge-type dwarf novae (e.g. figure 89 of \cite{Pdot4}).
Helped by the new finding and interpretation for the period variation
of the superhumps in the growing stage 
(\cite{osa13v344lyrv1504cyg}; \cite{kat13qfromstageA}), however,
we can now interpret that we observed unusually long-lasting
stage A superhumps which switched to stage B superhumps
in the middle of the rebrightening (see \cite{Pdot} for 
the description of stages A--C; while stages A--C usually
refer to the superhumps during the main superoutburst, the present
stage A and B superhumps were recorded during the rebrightening).
The growth of the superhump amplitude
during the rebrightening indicates that these superhumps are not
a result of the remaining eccentricity but reflect the newly
excited eccentric instability.

\begin{figure}
  \begin{center}
%    \FigureFile(75mm,90mm){j1222bshpdm.eps}
    \FigureFile(75mm,90mm){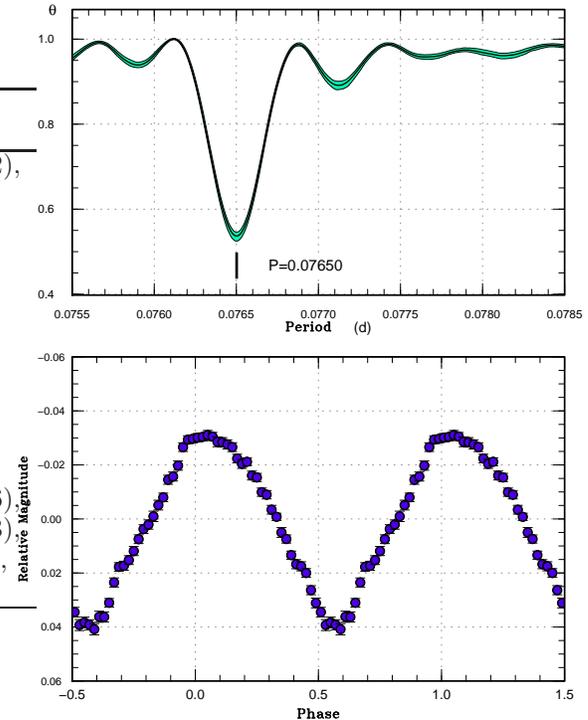}
  \end{center}
  \caption{Stage B superhumps in SSS J122221.
  (Upper): PDM analysis.  We analyzed 100 samples which randomly
  contain 50\% of observations, and performed PDM analysis for
  these samples.  The bootstrap result is shown as a form of 90\%
  confidence intervals in the resultant $\theta$ statistics.
  (Lower): Phase-averaged profile.}
  \label{fig:j1222bshpdm}
\end{figure}

   Stage A superhumps lasted at least for 150 cycles (it may be
even closer to 200 cycles if we assume the start of stage B
at the $O-C$ maximum).  We have determined the period of
stage A superhumps to be 0.07721(1)~d with the PDM analysis
for the segment $E \le 136$, where $E$ is the cycle count since 
BJD 2456304.808.
The period of stage B superhumps ($203 \le E \le 362$) was
0.07649(1)~d.  The profile of stage B superhumps is shown
in figure \ref{fig:j1222bshpdm}.  Note that the amplitude
is much smaller than in ordinary SU UMa-type dwarf novae,
suggesting the very small tidal torque.
The period derivative $P_{\rm dot} = \dot{P}/P$
of the stage B superhumps was almost zero
[$-1.1(7) \times 10^{-5}$].
There was no clear evidence for a stage B--C transition
as in most of WZ Sge-type dwarf novae \citep{Pdot}.

\subsection{Post-Superoutburst Superhumps}\label{sec:postsh}

   After fading from the rebrightening, the object showed
a longer period than the stage B superhumps.
The mean period was 0.07670(3)~d ($442 \le E \le 562$; 
We should note that we have no information whether there
was a phase jump between $E=362$ and $E=416$, and there may
be one cycle ambiguity in the cycle count).
Although individual times of maxima were not well determined
due to the faintness of the object, a PDM analysis of the later
post-rebrightening period (BJD 2456348--2456369) yielded
a period of 0.07676(4)~d.  We adopted an averaged value
0.07673(3)~d of the two methods.
This long superhump period
appears to be consistent with late-stage superhumps in some
WZ Sge-type dwarf novae (\cite{kat08wzsgelateSH}; \cite{Pdot2};
\cite{Pdot3}).

   According to the CRTS SSS data, the object remained $\sim$1 mag
brighter than in quiescence even after 82~d of the initial CRTS
detection (98~d after the initial MASTER detection).
Such a long-lasting fading tail is characteristic to
a WZ Sge-type dwarf nova.

\section{Discussion}\label{sec:discussion}

\subsection{Slow Evolution of Superhumps}\label{sec:shevolution}

   As shown in subsection \ref{sec:sh} it took 150--200 cycles
to fully evolve the superhumps.  Because of this very slow evolution
of superhumps, the stage A--B transition, which usually occurs
shortly after the outburst in ordinary SU UMa-type dwarf novae,
occurred in the middle of the rebrightening and produced
a peculiar $O-C$ pattern.  This phenomenon can be
naturally understood assuming a very small mass ratio and
very long growth time of the 3:1 resonance, which is expected
to be inversely proportional to $q^2$ \citep{lub91SHa}, 
and the superhump wave was confined to the 3:1 region for 
a long time due to the very small tidal effect.

   The growth time of superhumps in SSS J122221 was 5--7 times 
longer than typical short-period SU UMa-type dwarf novae 
(cf. \cite{Pdot}), suggesting that $q$ is 2--3 times smaller.
Assuming a typical $q$=0.10--0.15 for short-period 
($P_{\rm orb} \sim 0.06$~d) SU UMa-type dwarf novae,
we can expect $q \sim 0.05$ for this object.

\subsection{Estimation of the Mass Ratio from Precession Rates}
\label{sec:qestimate}

   It was very unfortunate that there was no chance for
time-resolved photometry during the initial outburst which 
is supposed to show early superhumps [double-wave modulations
in the early stage of WZ Sge-type outbursts, whose period is
very close to the orbital period (\cite{kat96alcom}; 
\cite{kat02wzsgeESH}).

   Despite the evidence for a rather high inclination from
spectroscopy, we could not detect any orbital signal in 
the post-rebrightening phase or the faint state before 
the rebrightening.  We therefore cannot directly apply 
the determination of $q$ using the stage A $\epsilon^*$ method 
\citep{kat13qfromstageA}.
We can put, however, a certain constraint.

   The outline of the method is as follows.
The precession rate of the elongated accretion disk,
which is the origin of the superhumps, can be determined
by a combination of the dynamical precession rate,
pressure effect, and the minor wave-wave interaction term
(\cite{lub92SH}; \cite{hir93SHperiod}).  The pressure
effect produces a retrograde precession, and reduces
the precession rate.  This effect can become negligible
when the superhump wave is still confined to
the radius of the 3:1 resonance (growing stage of superhumps,
observationally known as stage A superhumps) or when the disk is
cold (post-superoutburst stage).  During the growing stage of
superhumps, the precession rate is suggested to be
equal to the purely dynamical precession rate at the
radius of the 3:1 resonance \citep{osa13v344lyrv1504cyg}
and this has been confirmed by a comparison with the objects
with known mass ratios dynamically determined or determined by 
quiescent eclipses \citep{kat13qfromstageA}.

   The dynamical precession rate, $\omega_{\rm dyn}$
in the disk can be expressed by (see, \cite{hir90SHexcess}):
\begin{equation}
\label{equ:precession}
\omega_{\rm dyn}/\omega_{\rm orb} = Q(q) R(r),
\end{equation}
where $\omega_{\rm orb}$ and $r$ are the angular orbital frequency
and the dimensionless radius measured in units of the binary 
separation $A$.  The dependence on $q$ and $r$ are
\begin{equation}
\label{equ:qpart}
Q(q) = \frac{1}{2}\frac{q}{\sqrt{1+q}},
\end{equation}
and
\begin{equation}
\label{equ:rpart}
R(r) = \frac{1}{2}\frac{1}{\sqrt{r}} b_{3/2}^{(1)}(r),
\end{equation}
where
$\frac{1}{2}b_{s/2}^{(j)}$ is the Laplace coefficient
\begin{equation}
\label{equ:laplace}
\frac{1}{2}b_{s/2}^{(j)}(r)=\frac{1}{2\pi}\int_0^{2\pi}\frac{\cos(j\phi)d\phi}
{(1+r^2-2r\cos\phi)^{s/2}},
\end{equation}
This $\omega_{\rm dyn}/\omega_{\rm orb}$
is equivalent to the fractional superhump excess (in frequency)
$\epsilon^* \equiv 1-P_{\rm orb}/P_{\rm SH}$
and it is related to the conventional fractional superhump excess 
(in period) $\epsilon \equiv P_{\rm SH}/P_{\rm orb}-1$ 
by a relation $\epsilon^*=\epsilon/(1+\epsilon)$.

   We can now express fractional superhump excesses
(in frequency unit) of stage A superhumps and post-superoutburst
superhumps as follows:
\begin{equation}
\label{equ:epsstagea}
\epsilon^*({\rm stage A}) = Q(q) R(r_{\rm 3:1})
\end{equation}
and
\begin{equation}
\label{equ:epspost}
\epsilon^*({\rm post}) = Q(q) R(r_{\rm post}),
\end{equation}
where $r_{\rm 3:1}$ is the radius of the 3:1 resonance
\begin{equation}
\label{equ:radius31}
r_{3:1}=3^{(-2/3)}(1+q)^{-1/3},
\end{equation} 
$\epsilon^*({\rm post})$ and $r_{\rm post}$ are the fractional
superhump excess and disk radius immediately after
the outburst, respectively.  By solving equations
(\ref{equ:epsstagea}) and (\ref{equ:epspost}) simultaneously,
we can obtain the relation between $r_{\rm post}$ and $q$.
Since $r_{\rm post}$ is expected to be smaller than
$r_{\rm 3:1}$ and larger than the circularization radius
(0.20 $A$ for $q=0.05$ and 0.14 $A$ for $q=0.1$, \cite{lub75AD}),
we can put a constraint on the $q$ value.

   We can put a more stringent constraint on $q$ by using
experimentally derived $r_{\rm post}$ values.
According to \citep{kat13qfromstageA}, the disk radius of 
WZ Sge-type dwarf novae after the superoutburst (and associated
rebrightenings) was estimated to be 0.37--0.38 $A$ for
objects without rebrightenings and 0.30--0.32 $A$ for
objects with multiple rebrightenings.
If we assume a range of $0.30 \lesssim r_{\rm post} \lesssim 0.38 A$,
the $q$ range is $0.023 \lesssim q \lesssim 0.036$
(corresponding to a range of acceptable orbital period of
0.07612--0.07650~d).
This makes SSS J122221 one of the smallest $q$ known
in hydrogen-rich CVs (cf. \cite{pat11CVdistance}), and it is
the smallest $q$ estimated other than from fractional
superhump excess of stage B superhumps, which is not a good
estimator due to the strong pressure effect
(see \cite{kat13qfromstageA}).
Since SSS J122221 does not show helium enhancement
(e.g. \cite{mar13j1222atel4704}), this object is unlikely
a core-stripped compact binary similar to SBS 1108$+$574
(\cite{Pdot4}; \cite{car13sbs1108}; \cite{lit13sbs1108}).
Given the long $P_{\rm orb}$ and low $q$, this object is
a perfect candidate for a period bouncer.

\subsection{Possible Orbital Period}
\label{sec:porb}

\begin{figure}
  \begin{center}
%    \FigureFile(70mm,90mm){j122221porbpdm.eps}
    \FigureFile(70mm,90mm){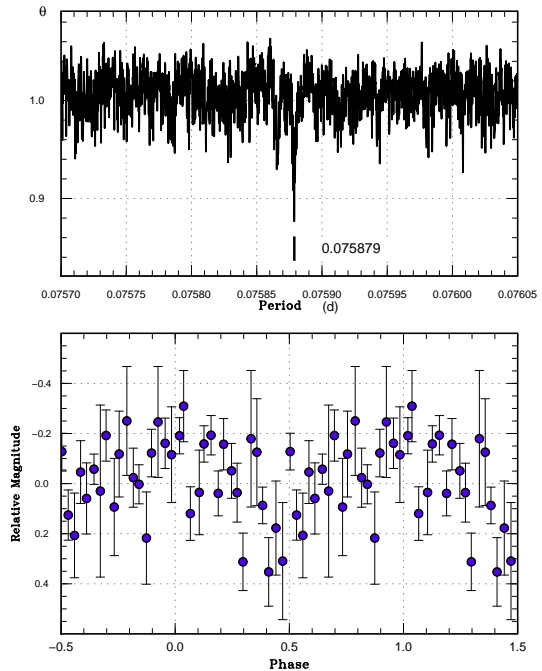}
  \end{center}
  \caption{Possible orbital variation in SSS J122221 in quiescence.
     (Upper): PDM analysis.
     (Lower): Phase-averaged profile.}
  \label{fig:j122221porbpdm}
\end{figure}

   An analysis of the CRTS data in quiescence
around this range of the orbital period yielded a possible
period of 0.075879(1)~d (figure \ref{fig:j122221porbpdm}).
This signal was present in an interval BJD 2453500--2454500,
but became weaker in an interval BJD 2454500--2455500.
This candidate photometric period may not be stable and 
needs to be tested by further observations.
If this period is the orbital period, it corresponds to 
$\epsilon^*$=0.017 for stage A superhumps and yields $q=0.045$.
The disk radius in the post-superoutburst phase
can be estimated to be 0.40 $A$.  Although this radius
is larger than in other WZ Sge-type dwarf novae,
a very small $q$ may be responsible.  As described in
\citet{osa95rzlmi} the disk radius at the end of
the superoutburst is a measure of the tidal strength and
\citet{osa95rzlmi} even assumed 0.42 $A$ for a low-$q$ object.
\citet{hel01eruma} proposed a similar idea of decoupling
between the thermal and tidal instability in low-$q$ systems.
We therefore consider the radius of 0.40 $A$ will not be
unexpectedly large.  The identification of the true orbital
period is wanted.

\subsection{Comparison with OT J184228.1$+$483742}
\label{sec:j1842comp}

   The peculiar pattern of a double superoutburst in SSS J122221
is not unprecedented.  OT J184228.1$+$483742 (=PNV J18422792$+$4837425;
hereafter OT J184228) showed a similar 
double superoutburst \citep{Pdot4}.  The initial superoutburst
in OT J184228 only consisted of early superhumps and (ordinary)
superhumps only grew during the second superoutburst (rebrightening).
Although the time-scales of the outbursts were longer in
SSS J122221, the overall pattern is very similar
(figure \ref{fig:j1222lc}).  \citet{Pdot4} considered that the
low $q$ and resulting long growth time of the 3:1 resonance
gives rise to this peculiar pattern of the outburst.
Although we used ``superoutburst'' for the first outburst
referring to its duration, the first outburst may bear 
characteristic of a prolonged precursor outburst whose 
long duration is sustained by the viscous depletion of 
the large amount of stored mass (as in the initial part of 
the ``case B'' outburst in \cite{osa03DNoutburst}).
\citet{kat13qfromstageA} also suggested $q$=0.042(3) 
for OT J184228 from stage A superhumps.
These two objects indeed appear to be very alike.

   We suggest that this kind of double superoutburst may be
characteristic to typical period bouncers, especially with
low $q$ and long $P_{\rm orb}$.  Until recently, a superoutburst 
with multiple rebrightenings such as in EG Cnc has been considered 
to a good indication of period bouncer (\cite{pat98egcnc};
\cite{pat11CVdistance}).  As the number of objects with
multiple rebrightenings has grown (ten objects at the time
of the writing), it has become evident
that many of these objects do not likely have $q$ as low as
in EG Cnc \citep{nak13j2112j2037}.\footnote{
  The $q$ value for EG Cnc was only estimated
  from $\epsilon$ for stage B superhumps, and this would suffer
  from an unknown pressure effect.}
We alternatively suggest that a double superoutburst as in
SSS J122221 or OT J184228 may be more typical to 
period bouncers, and that the case of EG Cnc may be exceptional.
Future measurement of $\epsilon^*$ for stage A superhumps
in EG Cnc will test this interpretation.

\medskip

We are grateful to the Catalina Real-time Transient Survey
team for making their real-time
detection of transient objects available to the public.
We thank Prof. Y. Osaki for comments.


\begin{thebibliography}{}

\bibitem[Bailey(1979)]{bai79wzsge}
  Bailey, J.\ 1979, \mnras, 189, 41P

\bibitem[{Carter} et~al.(2013)]{car13sbs1108}
  {Carter}, P.~J., {et~al.}\ 2013, \mnras, 431, 372

\bibitem[{Drake} et~al.(2009)]{CRTS}
  {Drake}, A.~J., {et~al.}\ 2009, \apj, 696, 870

\bibitem[{Drake} et~al.(2013)]{dra13j1222atel4699}
  {Drake}, A.~J., {et~al.}\ 2013, \ATel, 4699

\bibitem[Fernie(1989)]{fer89error}
  Fernie, J.~D.\ 1989, \pasp, 101, 225

\bibitem[{G{\"a}nsicke} et~al.(2009)]{gan09SDSSCVs}
  {G{\"a}nsicke}, B.~T., {et~al.}\ 2009, \mnras, 397, 2170

\bibitem[Hellier(2001)]{hel01eruma}
  Hellier, C.\ 2001, \pasp, 113, 469

\bibitem[Hirose, Osaki(1990)]{hir90SHexcess}
  Hirose, M., \& Osaki, Y.\ 1990, \pasj, 42, 135

\bibitem[Hirose, Osaki(1993)]{hir93SHperiod}
  Hirose, M., \& Osaki, Y.\ 1993, \pasj, 45, 595

\bibitem[Kato(2002)]{kat02wzsgeESH}
  Kato, T.\ 2002, \pasj, 54, L11

\bibitem[{Kato} et~al.(2013)]{Pdot4}
  {Kato}, T., {et~al.}\ 2013, \pasj, 65, 23

\bibitem[{Kato} et~al.(2009)]{Pdot}
  {Kato}, T., {et~al.}\ 2009, \pasj, 61, S395

\bibitem[{Kato} et~al.(2012)]{Pdot3}
  {Kato}, T., {et~al.}\ 2012, \pasj, 64, 21

\bibitem[{Kato} et~al.(2008)]{kat08wzsgelateSH}
  {Kato}, T., {Maehara}, H., \& {Monard}, B.\ 2008, \pasj, 60, L23

\bibitem[{Kato} et~al.(2010)]{Pdot2}
  {Kato}, T., {et~al.}\ 2010, \pasj, 62, 1525

\bibitem[Kato et~al.(1996)]{kat96alcom}
  Kato, T., Nogami, D., Baba, H., Matsumoto, K., Arimoto, J., Tanabe, K., \&
  Ishikawa, K.\ 1996, \pasj, 48, L21

\bibitem[{Kato}, {Osaki}(2013)]{kat13qfromstageA}
  {Kato}, T., \& {Osaki}, Y.\ 2013, \pasj, \inpress\arxiv{1307.5588}

\bibitem[Kato et~al.(2001)]{kat01hvvir}
  Kato, T., Sekine, Y., \& Hirata, R.\ 2001, \pasj, 53, 1191

\bibitem[Kolb, Baraffe(1999)]{kol99CVperiodminimum}
  Kolb, U., \& Baraffe, I.\ 1999, \mnras, 309, 1034

\bibitem[{Kuulkers} et~al.(2013)]{kuu13j1222atel4716}
  {Kuulkers}, E., {Page}, K.~L., {Knigge}, C., {Marsh}, T.~R., {Osborne},
  J.~P., \& {Sivakoff}, G.~R.\ 2013, \ATel, 4716

\bibitem[{Levato} et~al.(2013)]{lev13j1222atel4700}
  {Levato}, H., {et~al.}\ 2013, \ATel, 4700

\bibitem[{Littlefield} et~al.(2013)]{lit13sbs1108}
  {Littlefield}, C., {et~al.}\ 2013, \aj, 145, 145

\bibitem[{Lubow}(1991)]{lub91SHa}
  {Lubow}, S.~H.\ 1991, \apj, 381, 259

\bibitem[{Lubow}(1992)]{lub92SH}
  {Lubow}, S.~H.\ 1992, \apj, 401, 317

\bibitem[{Lubow}, {Shu}(1975)]{lub75AD}
  {Lubow}, S.~H., \& {Shu}, F.~H.\ 1975, \apj, 198, 383

\bibitem[{Marsh} et~al.(2013)]{mar13j1222atel4704}
  {Marsh}, T., {Knigge}, C., {Pretorius}, R., {Miller-Jones}, J., {Koerding},
  E., {Sivakoff}, G., {Woudt}, P., \& {Warner}, B.\ 2013, \ATel, 4704

\bibitem[{Nakata} et~al.(2013)]{nak13j2112j2037}
  {Nakata}, C., {et~al.}\ 2013, \pasj, \submitted

\bibitem[{Neustroev}, {Sjoberg}(2013)]{neu13j1222atel4744}
  {Neustroev}, V., \& {Sjoberg}, G.\ 2013, \ATel, 4744

\bibitem[Osaki(1995)]{osa95rzlmi}
  Osaki, Y.\ 1995, \pasj, 47, L25

\bibitem[{Osaki}, {Kato}(2013)]{osa13v344lyrv1504cyg}
  {Osaki}, Y., \& {Kato}, T.\ 2013, \pasj, \inpress\arxiv{1305.5877}

\bibitem[Osaki, Meyer(2003)]{osa03DNoutburst}
  Osaki, Y., \& Meyer, F.\ 2003, \aap, 401, 325

\bibitem[{Patterson}(2011)]{pat11CVdistance}
  {Patterson}, J.\ 2011, \mnras, 411, 2695

\bibitem[Patterson et~al.(1998)]{pat98egcnc}
  Patterson, J., {et~al.}\ 1998, \pasp, 110, 1290

\bibitem[{Pojma\'nski}(2002)]{ASAS3}
  {Pojma\'nski}, G.\ 2002, \actaa, 52, 397

\bibitem[Stellingwerf(1978)]{PDM}
  Stellingwerf, R.~F.\ 1978, \apj, 224, 953

\bibitem[Warner(1995)]{war95book}
  Warner, B.\ 1995, Cataclysmic Variable Stars (\PublisherCambridge)

\end{thebibliography}
\end{document}